\documentclass{pasa}

\jid{PASA}
\doi{10.1017/pas.\the\year.xxx}
\jyear{\the\year}

\usepackage[authoryear]{natbib}
\bibpunct{(}{)}{;}{a}{}{,}
\usepackage{aas_macros}
\usepackage{hyperref} 
\hypersetup{colorlinks,citecolor=blue,linkcolor=blue,urlcolor=blue}
\usepackage{gensymb}

\def\swift{{\em Swift}}
\def\fermi{{\em Fermi}}

\def\mJyPerBeam{{\rm mJy\,beam^{-1}}}

\makeatletter
\newcounter{MWAaffil}
\newcommand{\MWAaffil}[1]{%
  \ifcsname MWAaffil@#1\endcsname
  \else
    \stepcounter{MWAaffil}%
    \expandafter\xdef\csname MWAaffil@#1\endcsname{\theMWAaffil}%
  \fi
  \csname MWAaffil@#1\endcsname}
\makeatother

\def\builders{
and A.~P.~Beardsley$^{\MWAaffil{ASU}}$
and D.~Emrich$^{\MWAaffil{CurtinCIRA}}$
and T.~M.~O.~Franzen$^{\MWAaffil{Curtin}}$
and L.~Horsley$^{\MWAaffil{CurtinCIRA}}$
and M.~Johnston-Hollitt$^{\MWAaffil{Curtin}}$
and D.~L.~Kaplan$^{\MWAaffil{UWisc}}$
and D.~Kenney$^{\MWAaffil{Curtin}}$
and M.~F.~Morales$^{\MWAaffil{UW}}$
and D.~Pallot$^{\MWAaffil{UWA}}$
and K.~Steele$^{\MWAaffil{CurtinCIRA}}$
and S.~J.~Tingay$^{\MWAaffil{Curtin}}$
and C.~M.~Trott$^{\MWAaffil{Curtin},\MWAaffil{CAASTRO}}$
and M.~Walker$^{\MWAaffil{CurtinCIRA}}$
and R.~B.~Wayth$^{\MWAaffil{Curtin},\MWAaffil{CAASTRO}}$
and C.~Wu$^{\MWAaffil{UWA}}$
}
\def\affils{
\affil{$^{\MWAaffil{Curtin}}$International Centre for Radio Astronomy Research, Curtin University, Bentley, WA 6102, Australia}
\affil{$^{\MWAaffil{CurtinCIRA}}$Curtin Institute of Radio Astronomy, Curtin University, GPO Box U1987, Perth WA 6845}
\affil{$^{\MWAaffil{CAASTRO}}$ARC Centre of Excellence for All-sky Astrophysics (CAASTRO)}
\affil{$^{\MWAaffil{Astron}}$ASTRON, the Netherlands Institute for Radio Astronomy, Postbus 2, NL-7990 AA Dwingeloo, the Netherlands}
\affil{$^{\MWAaffil{APIA}}$Anton Pannekoek Institute for Astronomy, University of Amsterdam, Postbus 94249, NL-1090 GE Amsterdam, the Netherlands}
\affil{$^{\MWAaffil{CSIRO}}$CSIRO Astronomy and Space Science, PO Box 76, Epping, NSW 1710, Australia}
\affil{$^{\MWAaffil{USydney}}$Sydney Institute for Astronomy, School of Physics, The University of Sydney, NSW 2006, Australia}
\affil{$^{\MWAaffil{ASU}}$School of Earth and Space Exploration, Arizona State University, Tempe, AZ 85287, USA}
\affil{$^{\MWAaffil{UWisc}}$Department of Physics, University of Wisconsin--Milwaukee, Milwaukee, WI 53201, USA}
\affil{$^{\MWAaffil{UW}}$Department of Physics, University of Washington, Seattle, WA 98195, USA}
\affil{$^{\MWAaffil{UWA}}$International Centre for Radio Astronomy Research, University of Western Australia, Crawley 6009, Australia}
}

\title[MWA triggered observations]{A VOEvent based automatic trigger system for the Murchison Widefield Array}
\author[Hancock et al.]{P. J. Hancock$^{\MWAaffil{Curtin}}$\thanks{email: Paul.Hancock@curtin.edu.au}
and G. E. Anderson$^{\MWAaffil{Curtin}}$ 
and A. Williams$^{\MWAaffil{CurtinCIRA}}$
and M. Sokolowski$^{\MWAaffil{Curtin}}$
and S. E. Tremblay$^{\MWAaffil{Curtin},\MWAaffil{CAASTRO}}$
and A. Rowlinson$^{\MWAaffil{Astron},\MWAaffil{APIA}}$
and B. Crosse$^{\MWAaffil{Curtin}}$
and B. W. Meyers$^{\MWAaffil{Curtin},\MWAaffil{CAASTRO},\MWAaffil{CSIRO}}$
and C. R. Lynch$^{\MWAaffil{Curtin},\MWAaffil{USydney}}$
and A. Zic$^{\MWAaffil{USydney}}$
\builders{}\\
\affils{}
}

\begin{document}

\begin{abstract}
The Murchison Widefield Array (MWA) is an electronically steered low frequency ($<300$\,MHz) radio interferometer, with a `slew' time less than 8\,seconds.
Low frequency ($\sim 100$\,MHz) radio telescopes are ideally suited for rapid-response follow-up of transients due to their large field of view, the inverted spectrum of coherent emission, and the fact that the dispersion delay between a 1GHz and 100MHz pulse is on the order of $1-10$\,min for dispersion measures of $100-2000$\,pc/cm$^3$.
The MWA has previously been used to provide fast follow up for transient events including gamma-ray bursts, fast radio bursts, and gravitational waves, using systems that respond to gamma-ray coordinates network (GCN) packet-based notifications.
We describe a system for automatically triggering MWA observations of such events, based on VOEvent triggers, which is more flexible, capable, and accurate than previous systems.
The system can respond to external multi-messenger triggers, which makes it well-suited to searching for prompt coherent radio emission from gamma-ray bursts, the study of fast radio bursts and gravitational waves, single pulse studies of pulsars, and rapid follow-up of high-energy superflares from flare stars. 
The new triggering system has the capability to trigger observations in both the regular correlator mode (limited to $\geq 0.5$\,s integrations) or using the Voltage Capture System (VCS, $0.1$\,ms integration) of the MWA, and represents a new mode of operation for the MWA.
The upgraded standard correlator triggering capability has been in use since MWA observing semester 2018B (July-Dec 2018), and the VCS and buffered mode triggers will become available for observing in a future semester.

\end{abstract}
\maketitle%
\section{Introduction}%
\label{sec:intro}
The steady improvement in radio astronomy technology in recent decades has allowed for the deep study of the physics associated with transient astronomical events \citep{fender_radio_2011}, whether outbursting or explosive incoherent radio sources such as supernovae, gamma-ray bursts (GRBs) \citep{frail_radio_1997}, X-ray binaries, and tidal disruption events \citep{rees_tidal_1988}, or coherent sources such as fast radio bursts \citep[FRBs,][]{thornton_population_2013} and pulsars \citep{fender_transient_2015}.
Radio emission traces relativistic ejecta and unusual emission mechanisms, the observation of which allows astronomers to directly probe total energy budgets, magnetic fields and the properties and structure of the interstellar and intergalactic media.
However, the most extreme physics takes place at the very start of the transient event, such as the supernova shock `break-out' serendipitously detected in the X-ray band by \citet{soderberg_extremely_2008}, or the optical flash associated with the reverse shock emission from a GRB as seen by \citet{galama_effect_1999} and \citet{vestrand_bright_2014}.
In the case of FRBs, the transient event consists entirely of a single burst, with no afterglow yet detected \citep{williams_no_2016}.
In order to capture such short-lived associated emission, it is necessary for a variety of telescopes covering the entire electromagnetic spectrum to be capable of automatic and rapid follow-up of newly discovered astronomical transients.

Rapid-response follow-up of transients has previously been primarily conducted in the GHz regime, with telescopes such as a 12\,m dish based at the CSIRO Parkes Observatory, Australia \citep[1.4 GHz;][]{bannister_limits_2012}, the 26\,m dish located at the Mount Pleasant Radio Observatory, Australia,  \citep[2.3\,GHz;][]{palaniswamy_search_2014}, and
the Arcminute Microkelvin Imager (AMI) based at the Mullard Radio Astronomy Observatory (MRAO), UK \citep[14-18\,GHz;][]{staley_automated_2013,anderson_probing_2014,fender_prompt_2015,anderson_arcminute_2018}. 
The Australia Telescope Compact Array (ATCA) has also recently been equipped with a rapid-response systems \citep[e.g.][]{anderson_grb_2018}, further expanding the rapid-response frequency coverage in the GHz range from $1-20$\,GHz\footnote{https://www.narrabri.atnf.csiro.au/observing/users\_guide/html/\\chunked/ch02s05.html}.

While the first rapid-response radio experiments were conducted at 151\,MHz with the Cambridge Low Frequency Synthesis Telescope based at the MRAO \citep{green_search_1995,dessenne_searches_1996},
such programs in the MHz domain are only now resurfacing with the construction of the new generation of low frequency radio telescopes in preparation for the Square Kilometre Array (SKA). 
For example, the first station of the Long Wavelength Array \citep[LWA1, 10-88\,MHz;][]{taylor12lwa,ellingson2013lwa} has commissioned two rapid-response triggering modes with a 2 minute response time \citep{yancey15}. 
In October 2017, the High Band Array ($120-168$\,MHz) of the Low Frequency Array (LOFAR) enabled a rapid-response system capable of triggering within $3-5$\,min on transient alerts (referred to as the LOFAR Responsive Telescope\footnote{\href{https://www.astron.nl/radio-observatory/lofar-system-capabilities/responsive-telescope/responsive-telescope}{www.astron.nl/radio-observatory/lofar-system-capabilities/responsive-telescope/responsive-telescope}}). 
Meanwhile, the MWA, which operates  in the $80-300$\,MHz frequency range \citep{tingay_murchison_2013, wayth_phase_2018}, has been running a functional, yet somewhat limited rapid-response mode since December 2014 \citep{kaplan_deep_2015}.
There are also all-sky low frequency radio experiments that have been (or are capable of being) on-sky at the time of GRBs and gravitational wave events, including the LWA1 Prototype All Sky Imager \citep[LWA1-PASI;][]{obenberger14}, 
the Owens Vally Radio Observatory Long Wavelength Array \citep[OVRO-LWA, 27-84\,MHz;][]{anderson_simultaneous_2018,callister19}, and
the LOFAR Low Band Array ($10-90$\,MHz) Amsterdam ASTRON Radio Transient Facility and Analysis Centre \citep[AARTFAAC;][]{prasad14,prasad16}.

The MHz frequency range offers a number of benefits over the GHz regime due to a combination of intrinsic emission properties, propagation effects, and detector properties.
Non-thermal coherent emission typically has an inverted spectrum making such sources brighter at MHz frequencies.
The arrival time of pulsed signals are delayed with decreasing frequency due to dispersion caused by the ionized intergalactic and interstellar media \citep{taylor_pulsar_1993}, which means that MHz observations can be reliably triggered by gamma-ray, X-ray, or even GHz observations and still be on target before the signal arrives.
Radio interferometers in the MHz regime naturally have wide fields of view, are electronically steered, and often have large fractional bandwidths.
The MHz frequency range is the ideal observing band to search for prompt emission associated with transients.

\subsection{A rapid-response system for the MWA}
Since 2015, the MWA has been capable of receiving socket based alerts from the Gamma-ray Coordinates Network (GCN\footnote{gcn.gsfc.nasa.gov/gcn3\_circulars.html}).
Custom software \citep[based on that used for GRB triggering by][]{bannister_limits_2012} was built to analyse incoming GRB alerts from both the Niel Gehrels \swift{} Observatory \citep[hereafter referred to as \swift;][]{gehrels_swift_2004} Burst Alert Telescope \citep[BAT;][]{barthelmy_burst_2005} and the \fermi{} Gamma-ray Burst Monitor \citep[GBM;][]{meegan_fermi_2009}, and then automatically schedule 30 minutes of observations at the source position in the standard MWA observing mode.
Due to the limited amount of information available in the socket based GCN alerts, this system occasionally triggered on events that were not GRBs.
The \fermi{}-GBM and \swift{}-BAT can generate multiple alerts for the same burst, with updated positional information arriving at later times.
The typical 1 sigma error radius indicated in a \fermi{}-GBM trigger is $5-15$ degrees\footnote{\href{https://gcn.gsfc.nasa.gov/fermi.html}{gcn.gsfc.nasa.gov/fermi.html}} plus systematic uncertainties.
At $150$\,MHz the MWA has a field of view which is $15$ degrees radius at half power, which means that the \fermi{}-GBM positions would fall within this region approximately 2/3 of the time.
The MWA's socket based rapid-response system was unable to incorporate updated information from subsequent alerts on the same event into the observing schedule, resulting in the final GRB positions from \fermi-triggered events sometimes being located near the edge of, or even outside of the MWA's large field of view.
Nonetheless, the socket based alert system was successful at automatically observing GRBs.
Indeed, the observation of the short-duration GRB\,150424A set the most stringent upper limits (3\,Jy at 132\,MHz) on prompt radio emission from any type of GRB \citep{kaplan_deep_2015}.

We present the upgraded MWA rapid-response system, which now responds to transient alerts transmitted via the Virtual Observatory Event standard \citep[VOEvent;][]{Seaman_sky_2011}, that provide a machine readable format for the communication of astrophysical events.
All of the events that are distributed via the GCN are also distributed as VOEvents\footnote{gcn.gsfc.nasa.gov/voevent.html}.
The GCN distributed VOEvents contain more detailed information than that provided in the socket based alerts including: Moon and Sun angular distance, event position in multiple coordinate systems, spacecraft location, alternative classifications, and (probably most importantly) the probability of a given alert being (for example) a genuine GRB.
This increase in information makes it possible to trigger follow-up observations with more confidence, to reduce the fraction of false positives, and to update observations as new information becomes available.
The VOEvent format is an XML based format that can be easily interpreted by a variety of software such as the {\sc voevent-parse} python module \citep{staley_automated_2013, staley_voevent-parse_2014}.
We utilize the software packages provided by the `4 Pi Sky VOEvent Broker' \citep{staley_4_2016} and the Comet VOEvent client \citep{Swinbank_comet_2014} to parse and filter VOEvent transient alerts, enabling the automation of transient follow-up with the MWA.

The triggering system described here is able to observe in three modes.
The first mode is the regular correlator setup, which is used for most science observations with the MWA, and has a time resolution of $\geq 0.5$\,s.
The second mode is to capture voltages using the Voltage Capture System \citep[VCS;][]{Tremblay_high_2015} with a time resolution of $0.1$\,ms.
The third and final mode is a buffered capture mode, wherein the telescope is scheduled to capture voltages (with the VCS) to a ring buffer but only write to disk once a trigger is received.

In this paper we first review some science motivations for a triggered observing system on the MWA (\S\,\ref{sec:motivation}).
Next we describe the recently developed back-end service (\S\,\ref{sec:backend}) for the MWA telescope to enable fast rescheduling of the telescope, and the corresponding front-end (\S\,\ref{sec:frontend}) which will receive VOEvents and submit observing requests to the MWA.
We then describe the VOEvent filter that is in place to respond to \swift{} and \fermi{} GRBs (\S\,\ref{sec:grb}) as a case study.
We summarize and discuss future developments in \S\,\ref{sec:sumfut}.

\section{Science Motivations}
\label{sec:motivation}
This project was originally motivated by the desire to probe the very early-time low frequency radio emission from GRBs.
The overlap between short GRBs and gravitational wave events \citep[as demonstrated by the simultaneous detection of GW170817 and GRB 170817A;][]{abbott_multi-messenger_2017,abbott_gravitational_2017}, and the possibility of FRB-like signals being produced by these events \citep[for a review of these emission mechanisms see][]{rowlinson19}, provides further motivation.
The incorporation of voltage buffer triggers has made the triggering system particularly useful for the study of FRBs and intermittent pulsars.
Additionally, the fact that \swift{} and other space telescopes have on-board catalogues of known flaring stars, some of which are expected to have associated radio flares, means that triggered observations of M-Dwarf flares is another immediate science motivation.
In lieu of a complete list of science applications, we discuss here just those that are being actively pursued by the MWA using the triggering service.
Further discussion of the science applications of the MWA can be found in \citet{bowman_science_2013} and Beardsley et al. (2019, in prep).

\subsection{GRB and gravitational wave follow-up}
The study of short-duration GRBs (short GRBs) is highly topical as they are linked with binary neutron star (BNS), or neutron star (NS) - black hole (BH) binary mergers, which are the main class of gravitational wave (GW) events known to have electromagnetic counterparts \citep[ e.g. GW170817;][]{abbott_multi-messenger_2017,abbott_gravitational_2017}. 
There are several theories that predict such mergers should produce prompt, coherent emission in the form of a pulse, perhaps similar to FRBs, whether due to magnetic braking as the magnetic fields of the NSs are synchronized to the binary rotation before the merger \citep{hansen01,lyutikov13}, persistent or pulsating pulsar-like emission from a short-lived ($<1000$\,s), highly magnetized, supramassive neutron star remnant \citep{totani_cosmological_2013,rowlinson_signatures_2013, metzger_millisecond_2017}, or the collapse of said supramassive neutron star into a black hole \citep{falcke_fast_2014, zhang_possible_2014}. 
The detection of such prompt emission associated with the existence and/or collapse of a supramassive neutron star would allow us to constrain the equation-of-state of nuclear matter \citep[e.g.][]{lattimer_nuclear_2012,lasky_nuclear_2014}. 

Both short GRBs and GW events are therefore exciting targets for rapid-response observations with MWA.
The MWA has previously been involved in performing rapid-response triggered observations of both \swift{}- and \fermi{}-detected GRBs \citep[e.g. GRB 150424A;][]{kaplan_deep_2015}, as well as in the multi-wavelength follow-up of GW events \citep[e.g. GW170817;][]{abbott_multi-messenger_2017, andreoni_follow_2017}.
As the MWA is electronically steered, its rapid-response mode is capable of automatically repointing the telescope within 14\,s of receiving a transient alert (see \S\,\ref{sec:latency}).
This means that the MWA could be on target in time to detect any prompt emission associated with a BNS merger. 
In fact, \citet{rowlinson19} made predictions for such prompt emission associated with short GRBs over a wide range of redshifts and showed that MWA is very competitive for detecting such signals at low radio frequencies, particularly at the earliest timescales (seconds to minutes post-burst). 

Another benefit of the MWA for follow-up observations is its large field of view. While \swift{} GRBs are usually localised to within a few arcminutes, \fermi{} GRBs can have positional errors of up to tens of degrees, which can be well encompassed by the MWA's large field of view (see \S\,\ref{sec:grb} for further details).
In the case of GW events, the triggers from aLIGO/Virgo do not report a single pointing direction, but a probability map that spans many thousands of square degrees.
This large area of sky needs to be surveyed quickly in order to catch any prompt emission.
Again, the MWA's wide field of view makes is possible to quickly cover these large regions, and the automated rapid-response system described here allows for such a tiling of observations \citep[see][]{kaplan_strategies_2016}.

\begin{figure}
    \centering
    \includegraphics[width=1.05\linewidth]{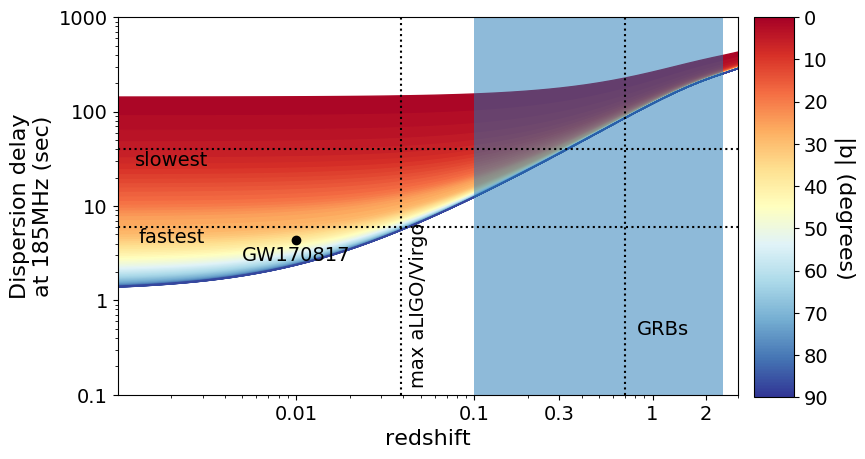}
    \caption{The dispersion delay at $185$\,MHz due to the intergalactic and interstellar medium as a function of redshift and absolute Galactic Latitude.
    The Galactic contribution is calculated from the model of \citet{yao_new_2017}, while the intergalactic contribution is computed from the model of \citet{inoue_probing_2004}.
    The redshift range of short GRBs is indicated in the shaded region, with the vertical dashed line representing the average redshift of 0.7.
    The expected dispersion delay for GW170817 and the horizon for detecting BNS mergers with aLIGO/Virgo during the O3 season are also indicated.
    The two horizontal lines indicate the fastest and slowest response times for the MWA, which are discussed in \S\,\ref{sec:latency}.
    }
    \label{fig:dm}
\end{figure}

Another consideration for both GRBs and GW events is the dispersion delay in the arrival time of any prompt emission at low frequencies, and whether MWA can be on-source in time to detect any such signals that may have been emitted at the time of the merger.  
Short GRBs have been detected by \swift{} at redshifts of between $0.1-2.5$ with an average of $z=0.7$ \citep{rowlinson_signatures_2013}.
\citet{yao_new_2017} model the Galactic contribution of dispersion measure as a function of sky position, while \citep{inoue_probing_2004} provide a model of dispersion measure from the intergalactic medium as a function of redshift.
Combining the Galactic and extra-galactic components with the observed redshift range for GRBs we compute that the arrival time of a prompt signal associated with the event would be dispersion delayed $12-404$\,s at $185$\,MHz, and take $4-132$\,s to cross the MWA's $30.72$\,MHz bandwidth \citep{taylor_pulsar_1993}.
However, the current sensitivity horizon for BNS events during the aLIGO/Virgo O3 run is 170\,Mpc so the expected dispersion delay for 
GW events can be much lower than for GRBs at high Galactic Latitude. 
The expected dispersion delay at 185\,MHz as a function of redshift is shown in Figure~\ref{fig:dm}, with the range Galactic dispersion measure contribution shown in colour.
The fastest and slowest reaction time of the MWA are indicated by horizontal dashed lines (see \S\,\ref{sec:latency}). 
We also show the range in redshift for a population of short GRBs, the expected dispersion delay for the BNS merger GW170817, as well as the horizon limit for the aLIGO/Virgo O3 observing run.  
Figure~\ref{fig:dm} therefore shows that the MWA will likely be on-target in time to observe prompt radio emission associated with most GRBs, however, is unlikely to respond fast enough to GW events, even if the aLIGO/Virgo alerts were instantaneous \citep[see][for MWA triggering predictions on negative latency GW alerts]{james_using_2019}.

An additional consideration is the effect of time sampling on the signal detection. 
The new ability of MWA to trigger higher temporal and spectral resolution observations using the VCS (see \S\,\ref{sec:vcs}) will increase the sensitivity to millisecond-duration pulses by at least an order of magnitude due to such pulses no-longer being smeared out over the coarse $0.5$\,s sampling of the standard correlator \citep[][shows the MWA VCS to be the most competitive triggering instrument for probing prompt, coherent radio emission from binary mergers]{rowlinson19}.
However, there are also deleterious effects of multipath scattering along the line of sight that act in opposition to this sensitivity improvement, and are especially potent at low frequencies since the pulse broadening time scale, $\tau$, is strongly frequency dependent, where typically $\tau\propto\nu^{-4}$ \citep[e.g.][]{geyer_scattering_2017, krishnakumar_multi-frequency_2017, bansal_scattering_2019, kirsten_probing_2019}.
The voltage data will be sensitive to this effect, and would allow us to resolve and place constraints on the scatter broadening of such pulses.
It is not expected that the pulses will be broadened so extremely (i.e. to beyond the 0.5\,s correlated observation time sampling) to entirely mitigate the effective sensitivity gained from acquiring the high time resolution time series data.
Therefore, the estimated order of magnitude improved sensitivity from capturing voltage data is a reasonably optimistic scenario.

\subsection{FRB observations}
\label{sec:frb}

To date, there are no reported detections of FRBs at frequencies lower than 400 MHz, even though a number of research groups have employed different techniques across a number of telescopes to search for them \citep[e.g.][]{coenen_lofar_2014,tingay_search_2015,rowlinson_limits_2016, chawla_search_2017}. Detecting the low frequency radio emission from FRBs, if it exists, would give unique insight into the emission energetics and would help to narrow down the progenitors from the large number of current candidates\footnote{\href{https://frbtheorycat.org}{frbtheorycat.org}} (as discussed below).
Similarly, the large fractional bandwidths inherent in low-frequency observations are of interest since the spectral modulation of FRBs has a high variance within the population \citep{shannon_dispersion-brightness_2018}. Despite this deficit, there is renewed hope at detecting FRBs at these frequencies with the Canadian Hydrogen Intensity Mapping Experiment (CHIME) detection of FRBs at the 400 MHz lower limit of their bandpass \citep{boyle_first_2018}.

The highly dispersed nature of these signals allows for a potentially more efficient use of low-frequency radio telescope time if one can trigger based on real-time detections of FRBs at higher frequencies. In the absence of such an automated triggering service, \citet{sokolowski_no_2018} have made use of shadowing observations to have the MWA co-observe with the Australian SKA Pathfinder \citep[ASKAP,][]{johnston_science_2008, hotan_australian_2014}.
In this shadowing setup, ASKAP observes the sky in a fly's-eye mode whilst recording baseband data \citep{james_performance_2019}, which is then processed off-line to search for FRBs, resulting in 20 new detections \citep{bannister_detection_2017,shannon_dispersion-brightness_2018}.
Simultaneously, the MWA observed an overlapping area of sky using the standard correlator mode ($10$\,kHz / $0.5$\,s resolution).
For each FRB detection by ASKAP, the MWA data were imaged and analysed for FRB emission.
The MWA did not detect any emission from the ASKAP detected FRBs in this mode of operation, providing insights into the spectral index of this enigmatic class of objects.
Namely, the non-detections are inconsistent with the mean spectral index of $\alpha=-1.8\pm 0.3$ that is measured for the brighest ASKAP detections.
Since pulse broadening cannot explain the non-detections with the MWA, this suggests a spectral turn over at frequencies above 200\,MHz, and plausible mechanisms are discussed, including: intrinsic spectral behaviour of the radiation process(es), free-free absorption, or caustic/scintillation induced amplification at higher frequencies \citep{sokolowski_no_2018}.

The current MWA-ASKAP shadowing setup has the MWA operating in the normal correlator mode, as the VCS data rate is too large to sustain for the multiple hours of observations that are carried out each day.
The ability for the MWA to operate in a buffered trigger mode, combined with a real-time detection system on ASKAP, would allow the MWA data to be captured at the best time and frequency resolution and only store those that are coincident with an FRB candidate.
This would increase the MWA's sensitivity to FRBs by at least an order of magnitude.

For FRBs with a DM as low as $100$\,pc/cm$^3$, which corresponds to a distance of $\sim 130$\,Mpc (depending on the line of sight, cf. FRB 171020; \citealt{shannon_dispersion-brightness_2018}), the time of arrival difference between detection at $1$\,GHz and pulse arrival at $100$\,MHz is just $40$ seconds.
This time delay is long enough that a real time FRB detection system on a 1\,GHz telescope can detect an event, generate a trigger, and pass it to a low frequency instrument such as the MWA, before the pulse arrives at MHz frequencies.
The time delay is short enough however, that an automated system is required both at the detection and follow-up stations.

\subsection{Pulsar observations}
\label{sec:pulsars}
Pulsars provide unparalleled Galactic laboratories to study astrophysical coherent emission processes.
While the emission from most pulsars is extremely regular, some pulsars exhibit irregular emission, such as sporadic emission \citep[e.g. giant pulses][]{meyers_spectral_2017}, or the switching of emission states \citep[e.g. ][]{kramer_periodically_2006,lorimer_radio_2012,young_apparent_2014}.
This sporadic emission occurs on time scales from seconds to months, and these pulsars pose a major challenge to understanding the underlying physics of the pulsar radio emission mechanism.
Simultaneous multi-frequency or contemporaneous high-energy and radio observations \citep[e.g. ][]{oronsaye_simultaneous_2015,meyers_spectral_2017,abdo_fermi_2010,hermsen_synchronous_2013,hermsen_discovery_2018} suggest that giant pulses and state-switching behavior are extremely broadband.

The vast majority of southern hemispheric pulsars are lacking low-frequency radio coverage, particularly for sporadic or intermittent pulsars.
Low-frequency observations with the MWA are thus very promising as they can reveal emission characteristics that are substantially different to those observed at higher frequencies.
In particular, sporadically emitting pulsars are inherently difficult to observe without regular monitoring and long dwell times, which is not currently feasible given the limited observing time available with the VCS (see \S\,\ref{sec:vcs}).
The newly-developed voltage buffer mode has been designed to mitigate these short-comings, especially the dwell time constraint (see \S\,\ref{sec:vcsbuffer} for details).

Both the MWA rapid-response triggering and buffering modes allow us to once again bypass the large VCS data recording rate, which is unsustainable for observations much longer than 1\,hr.
For example, telescopes that have regular, real-time pulsar monitoring programs, such as the recently upgraded Molonglo Observatory Synthesis Telescope \citep[UTMOST;][]{bailes_utmost_2017,2019arXivVenkatramanKrishnan}, can trigger the MWA VCS and/or the MWA buffering mode when, for example, intermittent pulsars are active, rather than relying on serendipity and potentially wasting valuable telescope observing time and resources. 
The recent detection of low-frequency emission from the intermittent pulsar J1107$-$5907 \citep{meyers_hunting_2018} provides an excellent demonstration of the MWA's ability to conduct such coordinated broadband observations. 
As it is, simultaneous broadband observations involving multiple telescopes can provide valuable insights into the pulsar emission mechanism, such as the spectral index distribution 
\citep[from both integrated profiles and single pulses, e.g.][]{meyers_spectral_2017,jankowski_spectral_2018} and single pulse energy distributions \citep[e.g.][]{burke-spolaor_high_2011,meyers_hunting_2018}, both of which are intimately tied to the emission physics.

\subsection{M-Dwarf flares}
\label{sec:mdwarf}
%% M-DWARFS
M-dwarf stars are known to produce frequent, powerful flares that are detectable across the entire electromagnetic spectrum. Simultaneous multi-wavelength observations of these flares have provided a window into the processes of plasma acceleration and heating within stellar atmospheres, which drive the flaring emission mechanisms in different wavebands (e.g. \citealp{2016ApJ...832..174O, 2010ApJ...721..785O, 2005ApJ...621..398O}, \citealp{fender_transient_2015}). 

At low radio frequencies, stellar flares are often coherent and highly polarized in nature.
Recently, the MWA has been successful in detecting faint, polarized flares from the well-known flare star UV Ceti \citep{2017ApJ...836L..30L}, uncovering a population of low frequency flares two orders of magnitude fainter than single-dish detections made before the 1980s (e.g. \citealp{1979MNRAS.187..405N,1974ApJ...190L.129S,1964Natur.203.1213L,1963Natur.199..991S}).

Some M-type stars exhibit gamma-ray or X-ray `super-flares', which are bright enough to trigger the \swift{}-BAT system. 
Recently, AMI triggered on a \swift{} gamma-ray super-flare from the nearby binary system DG CVn, detecting a bright ($\sim  100~\mathrm{mJy}$), incoherent flare at 15~GHz, 6 minutes after the gamma-ray detection, and an additional $90$\,mJy flare approximately 24 hours afterward \citep{fender_prompt_2015}.
These observations at radio frequencies were also accompanied by simultaneous observations using UV and optical facilities \citep{2016ApJ...832..174O}.
These multi-wavelength observations enabled a detailed analysis of the flare energetics, the relation of this powerful flare to lower-energy solar flares, and on the potential impact of such flares on the habitability of close-in planets around M-dwarfs \citep{2016ApJ...832..174O}.

It is currently unknown whether there is a low-frequency radio counterpart to the gamma and X-ray emission observed during M-dwarf super-flares. Recently, \citet{argiroffi_xraycme_2019} reported strong evidence of a coronal mass ejection for HR9024, where X-ray spectroscopy of the stellar flare was used to map plasma motions during the event. One possible source of low-frequency radio emission associated with powerful flares is from Type II bursts, produced during coronal mass ejections \citep{webb_reviewcme_2012,2018ApJ...862..113C,2006GMS...165..207G}. Rapid-response MWA observations of M-dwarf super-flares following triggers from high-energy facilities such as \textit{Swift} and the \textit{Monitor of All-sky X-ray Image (MAXI)}, would enable the potential detection of associated prompt, low-frequency flares, and any subsequent low-frequency emission associated with magnetospheric or coronal mass ejection-associated plasma motions.

\section{MWA triggering service}

The MWA rapid-response triggering system is divided into two parts: a back-end and a front-end.
The back-end is a web service that has been installed on an on-site server, forming part of the MWA Monitor and Control system.
This back-end system accepts requests from clients via the Internet.
An entirely separate front-end (which can be run externally or on-site) parses incoming VOEvents, makes decisions about when to trigger a new observation (or re-point an existing triggered observation), and calls the web service to schedule the observations.
Multiple front-ends responsible for monitoring and parsing different VOEvent streams and/or transient source types can be run in parallel.

Separating the science (what VOEvents to trigger on, and why) from the scheduling function lets the operations team handle the code that directly controls the telescope schedule, while allowing astronomers in the science project teams to write their own parsing code to decide which events to follow, and what observation and follow-up strategies to adopt.
The front-end system is able to accept pointing and frequency parameters as lists, which will then be iterated over to generate the final set of observing commands.
We now describe the two services.

\subsection{Back-end web service}
\label{sec:backend}

The back-end web service includes the following functions that are called by generating an HTTP request to a particular URL with a set of parameters:

\begin{itemize}
\item {\bf obslist} - When given a desired override duration, return a summary of all observations already in the schedule over that time period.

\item {\bf busy} - When given a science project ID code and a desired override duration (provided in seconds from the present), {\bf busy} returns `False' if that science project is authorized to override all of the observations already in the schedule over that time period.

\item {\bf triggerobs} - When supplied with a science project ID code and associated private key, {\bf trigger\_obs} will take a set of observational parameters (described later) and generate a set of observations using the standard MWA correlator.
If that science project is authorized to override all the observations already in the schedule over the requested time period, that period in the schedule is cleared and the requested observation/s are added to the schedule.

\item {\bf triggervcs} - like {\bf triggerobs}, but schedules observations with the Voltage Capture System \citep{Tremblay_high_2015} provided there is enough free disk space on the voltage capture servers.

\item {\bf triggerbuffer} - like {\bf triggervcs}, but for use when the VCS is currently in buffered mode (see \S\,\ref{sec:vcs}). {\bf triggerbuffer} does not accept observational parameters except for the observing duration, as the other parameters are already set when the telescope is put into the buffer mode.
Calling {\bf triggerbuffer} will cause the ring buffer to be drained and for VCS observations to continue as normal afterward.
\end{itemize}

When deciding if observations can be interrupted, the back-end software considers only the project codes for the existing and requested observations.
The prioritization of transient projects authorized to override active observations or another transient program is decided by, and at the discretion of 
the MWA Director\footnote{Including a consideration of the proposal scores assigned by the MWA Time Assignment Committee}, and encoded in a configuration file maintained by the operations team.
The only other additional constraint is that ongoing VCS observations can not currently be interrupted but this may change in the future.

Several input observational parameters can be specified in-order to fulfill different science requirements and to optimize the quality of the resulting data. 
The observational parameters can be given as lists so that observations will be scheduled that span multiple positions on the sky and multiple frequencies.
These inputs can include:
\begin{itemize}
\item {\bf (ra,dec) $\mid$ (alt, az) $\mid$ source}: A pointing direction or source name (from a limited local list of typical targets). Positions and names can be supplied as lists, and the back-end system will tile the sky accordingly.

\item {\bf avoidsun}: Whether to modify the given pointing direction/s to keep the desired target near the primary beam centre, but to minimise any contributed power from the Sun by placing it in a primary beam null (see \S\,\ref{sec:sun}).
This option has no effect if the Sun is below the horizon.

\item {\bf freqspec}: One or more frequency specifiers defining the (arbitrary) set of 24 coarse channels to record out of the coarse channels that define the MWA's $80-300$\,MHz frequency response. Each coarse channel is 1.28\,MHz wide making a contiguous bandwidth of $30.72$\,MHz, however, non-contiguous channel numbers may be specified.
Pointing directions and frequency specifiers are duplicated - for example, if two pointing directions are given, and three frequency specifiers, then each target direction will be observed at each of the three chosen frequency sets.

\item {\bf nobs, exptime}: The number of observations to schedule for each frequency/pointing combination, and the length of each observation.
By default, 15 consecutive observations of 120 seconds each are scheduled because the MWA analogue beamformers do not track sidereal motion during a single observation.

\item {\bf freqres, inttime}: The correlator averaging parameters to use - frequency resolution (currently 10, 20, or 40 kHz) and time resolution (currently 0.5, 1.0 or 2.0 seconds).

\item {\bf calibrator, calexptime}: Whether to schedule calibrator observation/s after the triggered source, and if so, what source to calibrate on, and how long the calibrator observation/s should be.
The user can also let the system choose a calibrator source automatically, by setting calibrator to be `True'.
If more than one frequency specifier was given, then the calibrator will be observed at each of the given frequency sets.

\end{itemize}

\subsubsection{Latency}
\label{sec:latency}
Fast response times are extremely important for a rapid-response triggering system, with lower latencies allowing astronomers to probe more exotic transient physics.
As the MWA is electronically steered, the repointing of the telescope using the rapid-response system is not limited by sky slew time, but rather in the automatic canceling and scheduling of observations.

The MWA observing schedule is stored in a set of database tables on a PostgreSQL\footnote{www.postgresql.org} server on-site, with start and stop times stored as the number of seconds since the Global Positioning System (GPS) epoch, referred to as `GPS seconds.'\footnote{\href{https://www.gps.gov/technical/icwg/\#is-gps-200}{IS-GPS-200}}

The MWA schedule works on a natural cadence of 8 seconds: all observations must start and stop on an integer multiple of eight GPS seconds.
This means that truncating an existing observation and inserting a new observation, will cause the new observation to begin on an $8$\,s boundary, leading to a natural latency of $0-8$\,s.
In practice, the Monitor and Control system gives the various components of the telescope time to prepare by sending their new configuration 4\,s ahead of the start of each observation.
This means that a running observation cannot have its stop time changed to a value less than $4$\,s in the future, and a new observation cannot be scheduled to start less than $4$\,s in the future. 
This effectively shifts the natural latency to be $4-12$\,s.
The triggering software and web back-end systems have a processing time of around 2\,s.
The Sun-avoidance check (\S\ref{sec:sun}, below) is not run when the Sun is below the horizon, but when it is run can add up to 10\,s of latency.
Choosing a calibrator automatically can take up to 8\,s, and changing the correlator mode to or from the VCS mode requires a dummy observation of 8\,s to be inserted into the schedule.
Including all the delays and cadences, the total latency period between the arrival of the transient alert notice and the start of a triggered observation is $6-40$\,s when triggering either the correlator or VCS rapid-response mode.
% 6s = 4s natural + 2s backend
% 40s = 12 natural + 10 avoid sun + 8sec calibrator + 8s mode change + 2s backend stuffs
For triggers of the voltage buffer, \textbf{no such latency exists} as all of the above mentioned setup is performed as the telescope enters the buffered mode, and the buffer can hold up to 150\,s of data, providing a `negative' latency for this mode.

\subsubsection{Sun avoidance}\label{sec:sun}
\begin{figure}
  \begin{center}
  \includegraphics[width=0.9\linewidth]{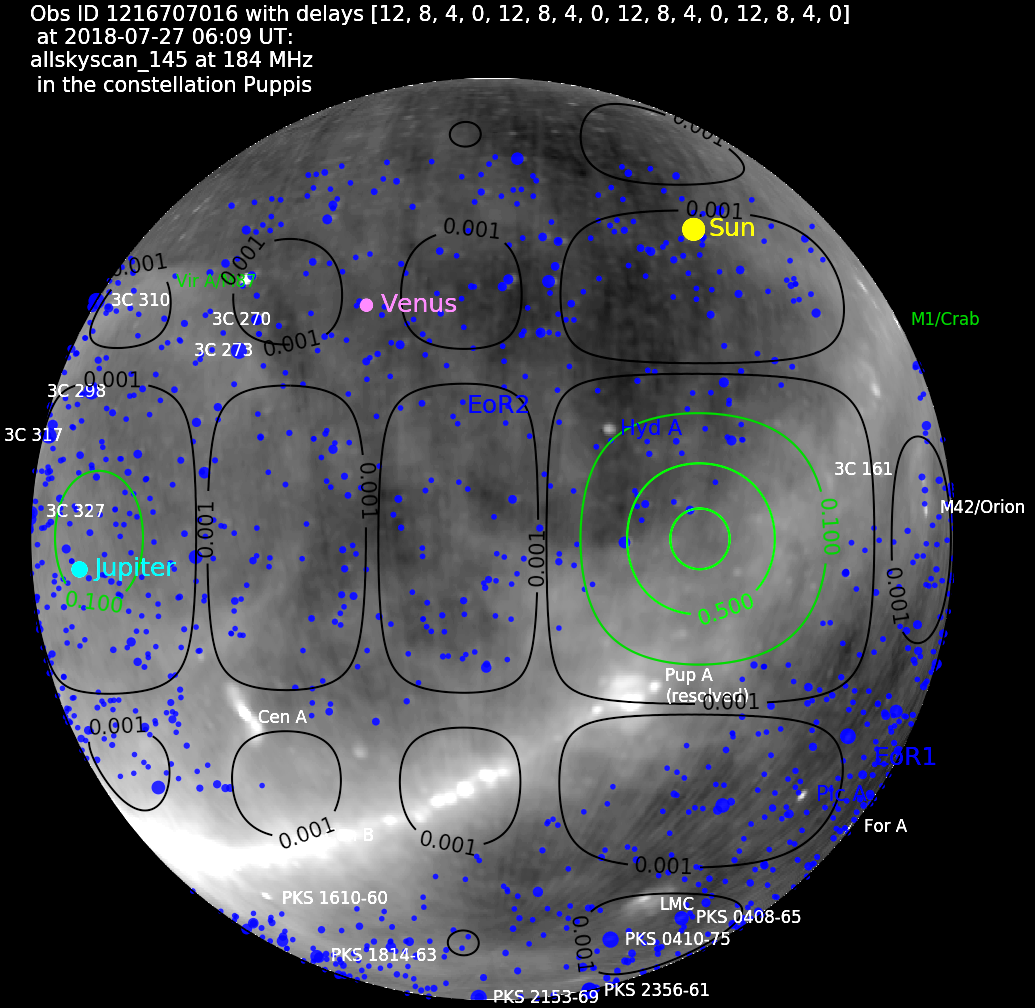}
  \caption{An all sky map showing: the radio continuum  from the Haslam map \citep[background grayscale;][]{haslam_408_1982}, the location of bright sources from the GLEAM catalogue \citep[blue circles;][]{hurley-walker_galactic_2017}, calibrators (named, white circles), solar system objects, and contours of the MWA primary beam normalized to the pointing direction (green and black).
During this observation the Sun was placed into a 0.1\% sidelobe, by the Sun avoidance code.
}
  \label{fig:pb}
  \end{center}
\end{figure}

Non-Solar daytime observations with the MWA are both possible and feasible.
For example, the rapid-response observation of GRB\,150424A occurred during the day \citep{kaplan_deep_2015}.
However, the main concern with daytime observations is the location of the Sun relative to the pointing direction and primary beam side-lobes.
Even if the Sun is located within a 1\% primary beam sidelobe, the power it contributes to the dataset creates artifacts across any resulting images which are difficult to remove.
Ideally, observations can be positioned such that the Sun is in or close to a primary beam null (Figure\,\ref{fig:pb}), thus minimizing its power contribution.
Since the primary beam of the MWA changes rapidly over the $30.72$\,MHz bandwidth, it may not be possible to place the Sun in a primary beam null over this entire band.

It is possible to adjust the requested pointing direction such that the sensitivity to the Sun is minimized, whilst the sensitivity in the direction of interest is maximized.
Therefore, if Sun avoidance is requested for alerts received during daytime, the system selects a pointing direction from the list of 197 MWA `grid-points' (pointing directions representing exact delays for all tile dipoles) in-order to minimize the contribution from the Sun.
The ratio, $r=B_{trg}/B_{sun}$, of the MWA primary beam sensitivity in the directions of the target ($B_{trg}$) and the Sun ($B_{sun}$) calculated at the centre of the observing band is used as a metric.
The values of the ratio $r$ for different `grid-points' can be as low as zero when the target is near the null of the primary beam ($B_{trg} \rightarrow 0$), and can approach infinity as the Sun approaches the primary beam null ($B_{sun} \rightarrow 0$).
Depending on the relative pointing directions of the target and Sun, the ratio $r$ can vary significantly (by even a few orders of magnitude) over the $30.72$\,MHz bandwidth. 
Therefore, a grid-point that optimizes $r$ at the central frequency may be sub-optimal at either end of the band.
Further revisions can be made during the data reduction level by sacrificing the upper- and lower-most parts of the band, which can result in an overall improvement in the quality and sensitivity of the images.

\begin{figure}
  \begin{center}
  \includegraphics[width=0.97\linewidth]{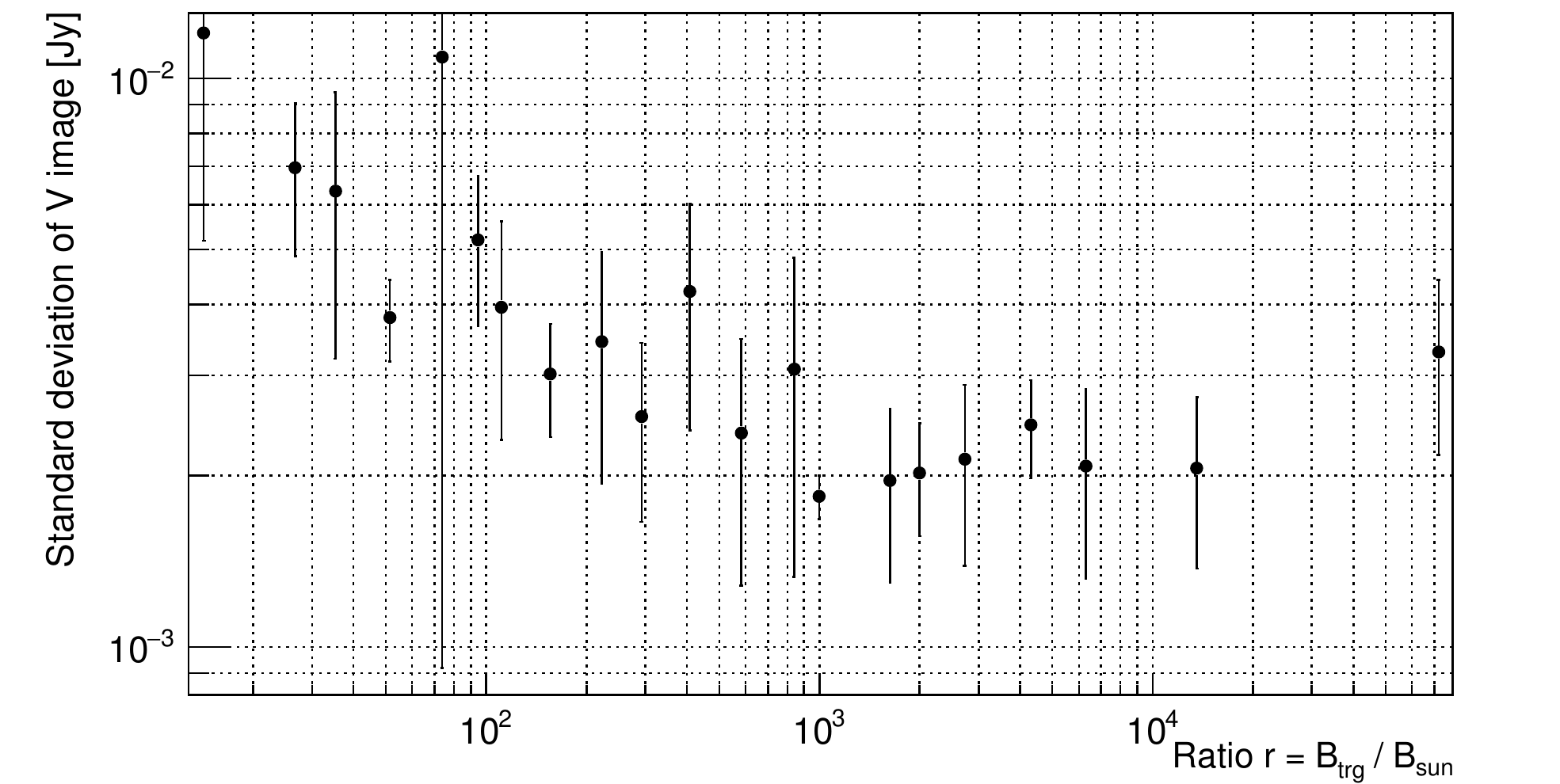}
  \caption{The dependence of the standard deviation of noise 
  in Stokes V images on the ratio ($r=B_{trg}/B_{sun}$) of the primary beam sensitivity in the directions of the target ($B_{trg}$) and the Sun ($B_{sun}$). The noise increases noticeably at ratios $r<1000$.}
  \label{fig_V_vs_ratio}
  \end{center}
\end{figure}

The optimal pointing is required to satisfy conditions: $r>1000$ and $B_{trg} \ge B_{min}$, where $B_{min}$ is the minimum acceptable beam response in the direction of the target ($B_{min} = 50$\%). 
This strategy is supported by Figure~\ref{fig_V_vs_ratio} which shows that the higher the ratio $r$, the better the properties of the image.
These initial criteria sometimes cannot be met, in which case we relax the criteria to allow $B_{min} \ge 10$\%.
This allows the software to quickly find a pointing direction with minimal response towards the Sun (i.e. maximizing the ratio $r$), which still provides at least $10$\% beam sensitivity towards a target.
Furthermore, the primary beam computations are one of the main contributions to the overall latency of the trigger system (see \S\ref{sec:latency}) so we opt not to decrease $B_{min}$ in stages, but instead jump directly to the minimum acceptable limit.

In order to test the Sun avoidance procedure, we performed daytime observations in all grid-point directions above elevation $50\degree$. 
The data quality was evaluated in terms of the standard deviation of the noise in Stokes V images ($\sigma_v$), since the image noise is not side-lobe confusion limited in circular polarization as it would be in total intensity.
The 25-second Stokes V images were of the quality expected for the night sky \citep[$\sigma_v \approx 2\,\mJyPerBeam$, based on predictions using the 2016 MWA beam model;][]{sokolowski_calibration_2017} for pointing directions where $r>1000$, whilst $\sigma_v$ noticeably increased in the pointing directions with $r<1000$ (Fig.~\ref{fig_V_vs_ratio}).

\subsubsection{Voltage Capture Mode}
\label{sec:vcs}
The MWA is capable of capturing high time and frequency resolution data using the VCS.
The VCS observing mode has a substantial data rate ($\sim 28\,$TB/hr), far greater than the rate at which it can be transported to the MWA archive in Perth in real-time so the data are initially stored on-site.
The on-site storage currently limits VCS observations to $\sim$ 90 minutes, so whilst it is possible to observe with the VCS, and possible to trigger this mode with the back-end mentioned in \S\,\ref{sec:backend}, additional constraints need to be placed on such observations.
To ensure that future VCS observations are not disrupted by a VCS trigger, a check is made to determine the current and expected disk use on-site at the MWA.
If the requested triggered observations would result in future observations failing due to insufficient disk space then the observing request will be rejected.

Each VCS observation stores data across 32 RAIDs simultaneously (2 per VCS server) so when a {\bf triggervcs} call is made, the software determines which RAID has the least free space and, based on that value, calculates how much time can be recorded.
The triggering system then checks what VCS observations are scheduled for the next 24 hours\footnote{For the current system. This is a configurable parameter.} and subtracts this scheduled observing time from the total recording time available.
If the duration of the requested trigger observation is shorter than the remaining time then a VCS trigger is allowed.

\subsubsection{Voltage Buffer Mode}
\label{sec:vcsbuffer}
There are some science cases where it is necessary to have the MWA actively follow the pointing direction of other telescopes (henceforth referred to as ``shadowing''), and the optimal observing strategy also requires the high time and frequency resolution provided by the VCS (e.g. ASKAP FRBs; see \S\,\ref{sec:frb}).
Unfortunately, the recording limit of $\sim 90$ minutes is not conducive to shadowing observations.
To address this, a new VCS observing mode was developed where the critically sampled (100\,$\mu$s/10\,kHz time and frequency resolution) tile voltages are stored in a ring buffer within the on-board memory of the VCS servers until a trigger is received, thereby mitigating the recording limit. 
In this buffer mode, the pointing direction and frequency selection is set prior to observing, and the telescope is collecting (but not always recording) data for the duration of this observation.
As such, normal telescope observations cannot continue when this mode is selected, unlike normal triggered observations that interrupt and override normal scheduled observations.

While operating in this mode, the voltage streams from all 128 tiles are stored in memory for up to 150 seconds before being discarded on a first-in-first-out basis, so at any given time, the last 150 seconds of data from the telescope is buffered in memory. When a trigger is received in this mode, the VCS software begins to write the buffered data to disk, while continuing to record new data to memory. This process runs in at least real time, thus after $\sim 2\text{--}3$ minutes the VCS servers have drained the memory buffers and will be operating as in a normal VCS observation, where the data are written directly to disk. Voltages will continue to be written to disk until the requested trigger stop time is reached. This mode is non-interrupting, in that the telescope is technically observing in Voltage Capture Mode (see \S\,\ref{sec:vcs}) for the scheduled duration even though potentially no data are being written to disk, and should therefore be thought of as a much more efficient, pre-scheduled VCS shadowing observation. Throughout the nominated observing time, the software will automatically re-point the telescope at a given duty cycle (typically 5--15 minutes) so that the target position is always within the tile primary beam.
It is possible to optimize these pointing directions in advance using the Sun avoidance mentioned in \S\,\ref{sec:sun}.

The voltage buffering functionality potentially allows for long-duration (i.e. $>90$ minutes) VCS observations where only those data containing an event are actually written to disk, thereby significantly reducing the overall data rate while providing a more versatile shadowing capability to the MWA.
Additionally, the buffer mode is critical for observations where the delay between high and low-frequency emission is not large (e.g. $\lesssim 40$ seconds).
For instance, sporadic emission from pulsars, where the dispersive delays can be many seconds, but still not long enough to process, send and receive triggers, re-point the MWA and begin VCS recording before the pulse arrives at $\sim 150$ MHz. 

An early prototype of this observing mode has been tested in exactly this circumstance, when UTMOST was used to trigger an MWA VCS recording of individual pulses from the intermittent pulsar J1107$-$5907 \citep{meyers_hunting_2018}.
In this case, the dispersion delay ($\rm DM = 40.75\,pc\,cm^{-3}$) between the two observing bands (835 and 154\,MHz for UTMOST and the MWA respectively) is only $\sim$8 seconds, shorter than the typical trigger latency, but easily within the voltage buffer capacity.

\subsection{VOEvent handler front end}
\label{sec:frontend}
The front end code uses the `4 Pi Sky VOEvent Broker' \citep{staley_4_2016} and the COMET VOEvent client \citep{Swinbank_comet_2014}.
The COMET client operates by receiving VOEvents and then making asynchronous calls to an external program.
We implement a simple queue to ensure that events are processed serially, and in the order in which they are received.
The script {\sc push\_voevent.py} is called by COMET and pushes the received VOEvent onto a handler service ({\sc voevent\_handler.py}) via a Remote Procedure Call (RPC).
The VOEvents are then passed to one or more registered handler functions in plugin libraries, which process the VOEvents using the {\sc voevent-parse} python module \citep{staley_automated_2013, staley_voevent-parse_2014} and submit observing requests if required.
The plugin event handlers return either True or False depending on whether an observing request was made.
{\sc voevent\_handler.py} will pass a VOEvent to all registered handlers stopping once a trigger request has been made or the list has been exhausted.
The module {\sc triggerservice.py} provides wrapper functions that abstracts the calls to the back-end web services: can\_interrupt\_now, obslist, triggerobs, and triggervcs.
The module {\sc handlers.py} provides a TriggerEvent object that holds information about triggered events, including a cache of VOEvents associated with the event. It also provides a high-level wrapper for triggering observations and also sends email notifications relating to the trigger.
Additionally, {\sc handlers.py} provides a wrapper function for retrieving the positional information from a VOEvent.
All the scripts noted thus far are provided to allow MWA science teams to write python scripts with a focus on parsing the information within the VOEvent without having to worry about the underlying data format or the HTTP requests that are used to trigger observations.

The front-end service is available at  \href{https://github.com/MWATelescope/mwa\_trigger}{github.com/MWATelescope/mwa\_trigger} under an Academic Free License (AFL 3.0), along with the VOEvent parser for the GRB follow up described below.

\section{Case study: GRB follow up}
\label{sec:grb}
\begin{figure}[h]
\centering
\includegraphics[width=0.98\linewidth]{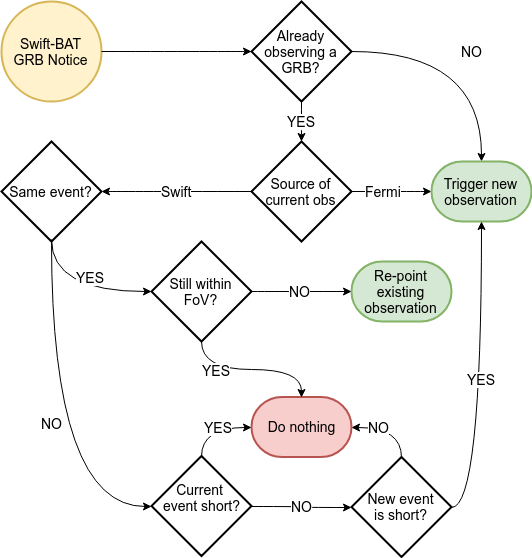}
\caption{Logical flow for receiving a \swift{} GRB alert.
The outcomes are either to trigger a new observation, update the current observation (green), or to not observe (red).
}
\label{fig:swift}
\end{figure}

\begin{figure}[h]
\centering
\includegraphics[width=0.98\linewidth]{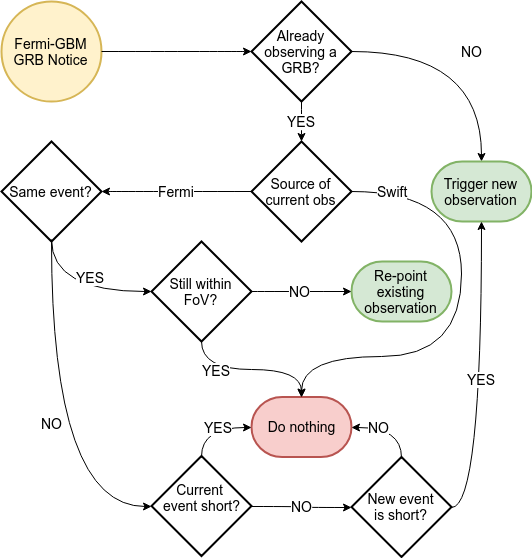}
\caption{Logical flow for receiving a \fermi{} GRB alert.
The outcomes are either to trigger a new observation, update the current observation (green), or to not observe (red).
}
\label{fig:fermi}
\end{figure}

In order to parse VOEvents of high-priority GRBs to the MWA back-end web service described in \S\ref{sec:backend}, we created the script {\sc GRB\_fermi\_swift.py}, which is capable of filtering GRBs detected by the \fermi{}-GBM and \swift{}-BAT instruments.
\swift{}-detected events are less common than \fermi{} as the BAT instrument can only observe one sixth of the sky when compared to the GBM, which monitors $\sim50\%$ of the sky at a time.
However, \swift{}-BAT events have a far better positional accuracy \citep[$1-4'$;][]{gehrels_swift_2004} compared to the \fermi-GBM \citep[with initial positional errors that are several 10s of degrees, with final position accuracy usually $<10$\,deg;][]{connaughton_localization_2015}.
We therefore give \swift-detected events priority over \fermi-detected events.
We also prioritize potential short GRBs over long GRBs, due to their association with gravitational wave events, and given that the majority of the emission mechanisms are expected to escape along the jet axis, they are less likely to be absorbed by the merger ejecta, which is expected to be more concentrated along the equatorial plane \citep{zhang_possible_2014}.
The high-level decision tree that {\sc GRB\_fermi\_swift.py} implements for parsing \swift{} and \fermi{} notices to MWA are outlined in Figures\,\ref{fig:swift}-\ref{fig:fermi}.

In Figures\,\ref{fig:swift}-\ref{fig:fermi}, the box `trigger new observation' represents a request for a new observation, which does not guarantee that an observation will take place.
The back-end triggering service makes the necessary checks to ensure that the requested observations are permitted to override the current observing program before they are scheduled.
We place an additional constraint that prevents any observing requests for a GRB that is below an elevation limit of $30^\circ$.
\fermi{} VOEvents also come with a MOST\_LIKELY parameter, which indicates the class of object that is most likely to have generated the \fermi{} alert.
We reject any alerts for which the likelihood of the event being a GRB is $<50\%$.
All \swift\ events are automatically compared to the Onboard Source Catalog (OSC).
If all known transients can be rejected, the \swift{} alert will automatically define the trigger as a real GRB, again allowing for further VOEvent filtering.
Note that the automatic \swift{} comparison to the OSC means that VOEvents can also be monitored for other source types, such as known X-ray binaries, flare stars and magnetars, potentially motivating other triggering programs.

For all alerts, we make the assumption that the most recent notice will have the most up-to-date information and so we will re-point to the newest position if it is not in the field of view\footnote{In practice we define a minimum angular offset (default is $10\deg$), and only new-positions that are more distant than this will cause a new trigger.} of the current observation.
The exception is that \fermi{} alerts arrive in three different flavours: flight position, ground position, and final position.
For a single event, \fermi{} can generate zero, one or more of each flavour of alert, and not always in the same order.
We ascribe a positional reliability hierarchy to the \fermi{} alerts with the highest to lowest reliability being: final $>$ ground $>$ flight.
Notices of the same flavor will cause a position update check, but those of lower reliability cannot update positions previously generated from a higher priority alert.
For example, a ground position notice will generate a re-pointing request if the best position was generated from a previous ground position or flight position alert but not if it was generated from a final position alert.

A given GRB can also be detected by both \swift{} and \fermi{}.
The Fermi-GBM triggers often arrive before the Swift-BAT triggers \citep[e.g. GRB 130427A;][]{anderson_probing_2014}.
Given the much better positional precision of \swift\ generated events, they are prioritized over the \fermi\ alerts. We therefore make no effort to cross-identify GRBs that are detected by both telescopes.
Such GRBs will always generate observing requests and will do so at the more precise \swift{} position.
While we prioritize the follow-up of short GRBs, it is not possible to determine the long/short nature of a GRB at the time of detection, the classification of which usually requires human inspection of the gamma-ray data.
We therefore only trigger on those \fermi\ events for which the integration time (the time in seconds it takes the gamma-ray signal to reach the triggering threshold of the instrument, which is different to the burst duration) is $\le 2.048$\,s. This $2$s is based on the classification of the short GRB class \citep{kouveliotou_identification_1993}, whilst the fractional quantity incorporates the resolution of the duration in \fermi{} notices.
We use the same cut-off for \swift\ events, allowing us to re-point at any new \swift{} triggers that are more likely to be a short GRB (see Figure~\ref{fig:swift}). 

\section{Summary and future development}
\label{sec:sumfut}
We have developed a VOEvent based rapid response system for the MWA telescope.
This development was motivated by the need to perform low-latency observations of GRBs detected by the \swift{} and \fermi{} telescopes, and to enable efficient use of the voltage capture system for the study of GRBs, FRBs, and intermittent pulsars.
The software runs in two parts: a back-end system that is part of the MWA Monitor and Control system, which exposes multiple web interfaces, and a front-end system that can be run anywhere, responds to VOEvents and interacts with the back-end system.
The front-end system, along with example VOEvent handlers, is available at \href{https://github.com/MWATelescope/mwa\_trigger}{github.com/MWATelescope/mwa\_trigger}.
Contributions, bug reports, and feature requests are encouraged.

The system is still being developed, and new features are being planned.
The latency of the back-end system is currently $6-40$\,sec.
A natural $4-12$\,sec latency exists as part of the scheduling cycle of the MWA, and this may be reduced as part of a future upgrade to the MWA.
The remaining $2-28$\,sec latency is due to: Sun-avoidance calculations, calibrator selection and scheduling, and correlator mode change operations.
The Sun is not the only bright source that causes consternation: bright radio sources in the primary beam side-lobes can also cause calibration issues during both day time and night time observations.
An extension of the Sun avoidance code is planned for night time observations, so as to place these difficult sources in the null of the primary beam at the central observing frequency.
Currently the back-end system inserts observations into the schedule by expunging and truncating existing observations, and there is not yet any concept of canceling a triggered observation.
In the future we plan to have the deleted observations cached so that if a trigger is canceled, the MWA schedule can be re-instated, reducing the impact on other observers.

This work represents a new mode of operation for the MWA.
The upgraded standard correlator triggering capability has been implemented since MWA observing semester 2018B, and the VCS and buffered mode triggers will become available for observing in a future semester.

\begin{acknowledgements}
\subsubsection*{People}
We thank the referee for their comments and suggestions, which improved the quality of this publication.
We thank Natasha Hurley-Walker, Freya North-Hickey and Jean-Pierre Macquart for their suggestions that led to the creation of Figure 1.
We thank the current and previous MWA Directors, Melanie Johnston-Hollitt and Randall Wayth, for providing director's time during the development and testing of this service.
We acknowledge the Wajarri Yamatji people as the traditional owners of the site on which the MWA operates.

\subsubsection*{Funding}
GEA is the recipient of an Australian Research Council Discovery Early Career Researcher Award (project number DE180100346) funded by the Australian Government.
BWM acknowledges the contribution of an Australian Government Research Training Program Scholarship in supporting this research.
SET and BWM acknowledge the Australian Research Council grant CE110001020 (CAASTRO).

\subsubsection*{Software}
The front end software makes use of COMET \citep{Swinbank_comet_2014} and {\sc voevent-parse} \citep{staley_4_2016} to process the incoming VOEvents.
Both the front-end and back-end software rely on the Astropy \citep{the_astropy_collaboration_astropy_2013,astropy_collaboration_astropy_2018}, numpy \citep{van_der_walt_numpy_2011}, and scipy \citep{Jones_scipy_2001} python modules.

\subsubsection*{Facilities}
This scientific work makes use of the Murchison Radio-astronomy Observatory, operated by CSIRO. We acknowledge the Wajarri Yamatji people as the traditional owners of the Observatory site. Support for the operation of the MWA is provided by the Australian Government (NCRIS), under a contract to Curtin University administered by Astronomy Australia Limited. We acknowledge the Pawsey Supercomputing Centre, which is supported by the Western Australian and Australian Governments.

This work made use of data supplied by the \swift{} satellite. \swift{}, launched in November 2004, is a NASA mission in partnership with the Italian Space Agency and the UK Space Agency. \swift{} is managed by NASA Goddard. Penn State University controls science and flight operations from the Mission Operations Center in University Park, Pennsylvania. Los Alamos National Laboratory provides gamma-ray imaging analysis.

\subsubsection*{Services}
This research has made use of NASA's Astrophysics Data System, and the online cosmology calculator tool\footnote{\href{http://www.astro.ucla.edu/~wright/CosmoCalc.html}{www.astro.ucla.edu/$\sim$wright/CosmoCalc.html}} \citet{wright_cosmology_2006}.

\end{acknowledgements}

%\nocite*{}
\bibliographystyle{pasa-mnras}
\bibliography{TriggeringPaper,MDwarf,manual_refs}
\end{document}